# Hydrogen dissociation catalyzed by carbon coated nickel nanoparticles: experiment and theory

A. Ye. Yermakov,[a] D. W. Boukhvalov,*[b] M. A. Uimin,[a] E. S. Lokteva,[c] A. V. Erokhin,[c] N. N. Schegoleva[c]

*Based on combination of experimental measurements and first-principles calculations we report a novel carbon-based catalytic material and describe significant acceleration of the hydrogenation of magnesium at room temperature in presence of nickel nanoparticles wrapped in multilayer graphene. Increase of the rate of magnesium hydrogenation in contrast to the mix of graphite and nickel nanoparticles evidences intrinsic catalytic properties of explored nanocomposites. Results of simulations demonstrate that the doping from metal substrate and the presence of Stone-Wales defects turn multilayer graphene from chemically inert to chemically active mode. The role of the size of nanoparticles and temperature are also discussed.*

## Introduction

Recent progress in carbocatalysis[1-4] when the catalysts are based on modified (doped or functionalized) graphenes demonstrates significant increasing of reaction rate and the number of the working cycles for various reactions such as oxygen reduction or hydration or oxidation of various organic species encourage us examine other type of imperfect graphenes (multilayer graphene with defects over nickel substrate)[5,6] as the catalysts. Hydrogen dissociation reaction has been chosen for a probe of catalytic properties of these materials due to importance for the further applications, simplicity in modelling and count of reaction yield.

Molecular hydrogen dissociation is necessary for the formation of metal hydrides, the most prospective materials for hydrogen storage due to its high hydrogen weight content, cost and abundance of used materials.[7,8] Especially attractive is the use of light metals for hydride formation, between them Mg is the lightest and inexpensive. Conventional way for improving the reaction rate (the increasing of the temperature and pressure of hydrogen gas) is ineffective: it takes at least several hours to prepare the $MgH_2$ hydride from pure magnesium even at 350–400 °C and hydrogen pressure of 3MPa.[9] Two main problems hinder the use of the magnesium as storage material for hydrogen: the high binding energy of hydrogen and slow diffusion of hydrogen through the $MgH_2$ layer that forms on the Mg surface and obstruct hydrogen uptake.[10] Ball milling and addition of catalysts were proposed to improve hydrogen uptake kinetics. The decrease of Mg grain size during ball milling causes an increase in the specific Mg surface area and shortens the diffusion pathways, but can also cause the decrease of hydrogen storage capacity.[11] Searching of the most efficient catalyst for magnesium hydrogenation is the subject of plenty experimental works (see Refs. 9 and 12 and references therein). Our recent experimental work[13] also demonstrates that $V_2O_5$ is the most efficient from studied compounds catalytic material for this reaction by the ratio of the weight and yield (see Fig. 1).

Ball milling of Mg in the presence of graphite[14,15] or nanotubes[16] causes the improvement of kinetics of hydrogen uptake; it is attributed to charge transfer between Mg and graphite or hindering oxygen back-diffusion to the Mg surface that improves hydrogen dissociative chemisorption. To address the problem of hydrogen binding energy Mg composites were used instead of pure Mg, such as Mg-C composite, doped with Pd.[17] The loading proceeds very fast, during a few seconds, while slow relaxation is seen thereafter. The authors explain such behavior by broad distribution of binding energies for Mg-C thin film, doped with Pd, namely from bulk $MgH_2$ (−0.38 eV/atom) levels and to levels as low as −0.3 eV/atom in a small part of the sample. However, the influence of Pd that can absorb huge amounts of hydrogen is not discussed.

Thus the use of metal-carbon nanocomposite as the catalyst for hydrogen uptake with Mg is the next step in the development of the catalytic materials for hydrogen decomposition. Ni@C composite contains graphene-like carbon as outer shell and the nickel core. Ni@C nanocomposites were prepared by melting in suspension state and evaporation of melted metal in a flow of inert gas containing light hydrocarbons. Carbon coated nickel nanoparticles consist of nickel core about 4-5 nm in size wrapped in a few layers of carbon (see Refs. 5, 6). Evaluation of atomic and electronic structure of these particles[6] demonstrates that carbon shell of nickel nanoparticles consists of a few stable graphene layers with significant amount of Stone-Wales[17] defects. Formations of these graphene layers (ten or less) over nickel substrate is similar to the formation of perfect graphene mono and few layers over transition metal substrate[19-21] and protect the transitional metal core for any interactions with environment within two years.[5] Previous theoretical calculations predict the enhancement of chemical activity of graphene comprising this type of defects[22] or doped by transition metal impurities.[23] The combination of these reasons with low cost and facility in production, long time stability at ambient conditions[6] and abundance of by-products (or wastes) motivate us examine the catalytic properties of carbon coated nickel nanoparticles in the chemical reaction with rather simple mechanism and high importance for the further applications. On the other hand hydrogen dissociation reaction is a feasible test for the examination and understanding of the catalytic properties of these materials.

[a] Prof. A. Ye. Yermakov, dr. M. A. Uimin
Institute of Metal Physics, Ural Branch of RAS,
620990 Yekaterinburg, Russian Federation
[b] Prof. D. W. Boukhvalov
School of Computational Studies
Korea Institute for Advanced Study (KIAS)
Seoul 130-722, Korea.
[c] Prof. E. S. Lokteva, A. V. Erokhin, dr. N. N. Schegoleva
Lomonosov Moscow State University, Chemistry Department,
119991 Moscow, Russian Federation.

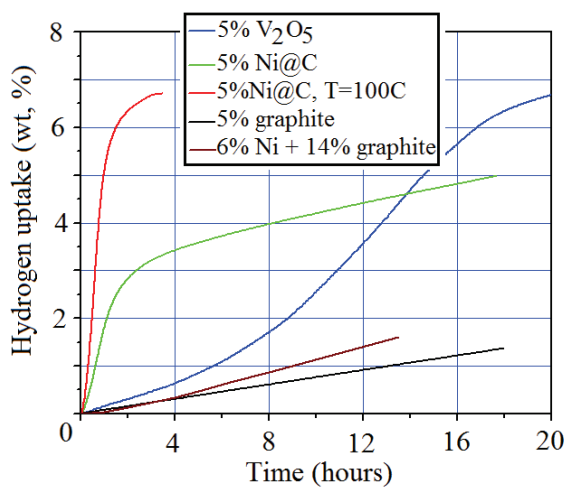

Figure 1. Hydrogen uptake in magnesium hydride as a function of milling time in hydrogen atmosphere and the type and percentage of catalyst ($V_2O_5$, graphite powder, Ni nanoparticles and graphite powder mix (all at room temperature) and carbon coated Ni nanoparticles (Ni@C) at room and 100°C temperature.

## Results and Discussion

Hydrogenation of Mg was realized by vibratory ball-milling set up using a steel container and balls in the hydrogen atmosphere in the presence of different catalysts (see experimental section). To control the hydrogen uptake by Mg the hydrogen pressure in container was measured in the process of mechanical treatment.[13]

The results of magnesium hydrogenation (Fig. 1) in the presence of different amount of the carbon coated nickel nanoparticles demonstrate significant increasing of magnesium hydrogenation rate in the presence of minimal (5% of magnesium and catalyst weight) amount of carbon coated nickel nanoparticles in contrast to the $V_2O_5$ catalyst. Reaction time decreases from 10 (in the presence of $V_2O_5$) to 2 hours (in the presence of Ni@C). Further increasing of the concentration of our nanoparticles to 20% provides subsequent increasing of magnesium hydrogenation rate. To examine of the role of carbon core and nickel shell we also have performed this reaction in the presence of the mixture of pure nickel powder and graphite and found that the efficiency of the this mixture is lower than that of carbon coated nickel nanoparticles. The contrast between the results for nickel nanoparticles with and without carbon shell evidences the crucial role of the carbon shell for the catalytic properties. It seems that Ni role is to modify carbon properties and provide tiny particulate substrate to form defected graphene layer on it. The absence of direct catalytic effect of Ni is in accordance with the data presented by Y.J. Chai at al.[24] where nanoporous Ni layer microencapsulated Mg showed a reversible hydriding-dehydriding process without formation of stable hydrides; obvious physisorption characteristic for porous Ni layer rather than hydrogen storage behaviour of Mg was observed for such system.

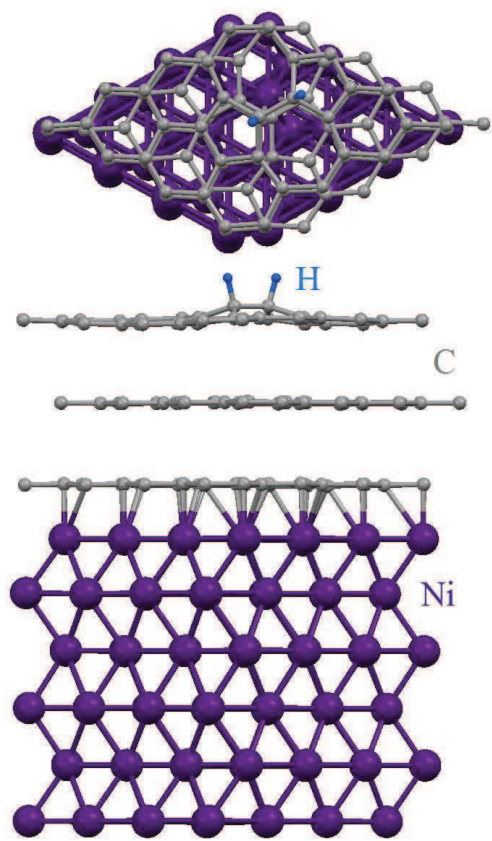

Figure 2. Top and side view of the optimized atomic structure of graphene trilayer over Ni scaffold with hydrogen pair chemisorbed on Stone-Wales defect.

For the modelling of hydrogen decomposition over Ni@C nanoparticles we have used mono or trilayer graphene 3×3 supercell containing 32 atoms in each layer free standing or over 6 layers of hcp-Ni which is imitate Ni(111) surface (see. Fig. 2). Based on the previous experimental and theoretical studies[6] we performed the calculations taking into account the presence of the number of Stone-Wales defects. To check the catalytic activity we examine different levels of the hydrogen coverage of graphene substrate (see Fig. 3).

Results of our calculation (Fig. 4) demonstrate significant decrease of the chemisorption energy of the first hydrogen molecule over graphene with Stone-Wales defects over nickel substrate in contrast to the chemically inert at room temperature pristine graphene.[25,26] The choice of the ball-milled graphite (without SW defects) as a catalyst does not lead to increasing the rate of Mg hydrogenation. The energy of chemisorption of hydrogen pair on graphene with SW defects on nickel carrier is close to zero that corresponds to the formation of metastable carbon–hydrogen bond. This temporary pair of hydrogen atoms could be formed during adsorption of molecular hydrogen and desorbed at room temperature. These results can be also feasible for the graphene produced by chemical vapour deposition on Ni or Cu (111) surface that also contains some amount of Stone-Wales defects.[19,20]

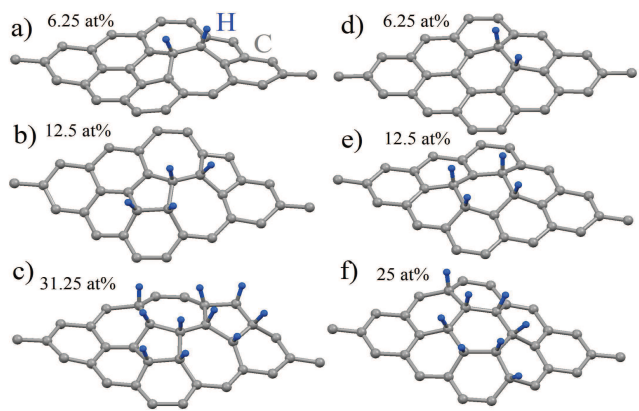

Figure 3. Optimized atomic structure of graphene monolayer with (a-c) and without (d-f) Stone-Wales defects at different level of one-side hydrogenation of surface.

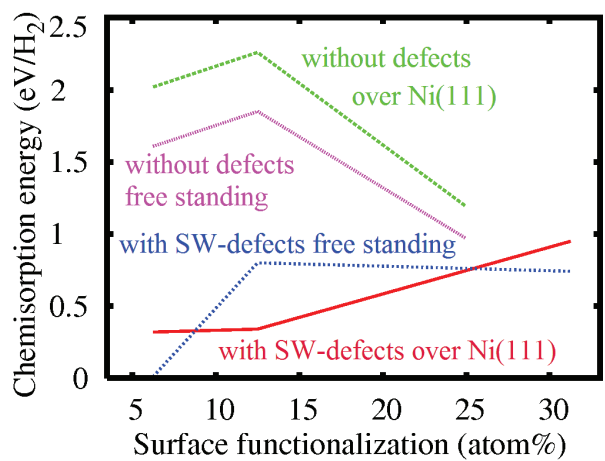

Figure 4. Chemisorption energies as a function of hydrogenation level for the first, second and third steps of hydrogen load (see Fig. 3) calculated for graphene trilayers with or without Stone-Wales (SW) defects and/or nickel substrate.

The values obtained in the calculations for the first and second pairs of hydrogen chemisorption on graphene with Stone-Wales defects (Fig. 3a,b) on Ni(111) scaffold is lower than the energy required for water evaporation, defined as the difference between the formation energies of water in the liquid and gaseous phases (43.98 kJ/mol[27] or 0.46 eV). Thus, considering the energetics alone, these calculations showed that the adsorption of molecular hydrogen and dissociation with further desorption of atomic hydrogen over defective graphene on metallic substrate can occur at room temperature. In order to estimate the connection between the calculated energy cost of the reactions and experimental results obtained at higher temperatures (100 and 140 ℃), we compared the calculated energies with the ones required for the deeply experimentally and theoretically explored case of monolayer graphene oxidation (about 1.4 eV[28] with the experimental results of graphene monolayer fast oxidation at 200 ℃ [26]). Discussed in refs. 26, 28 value of graphene oxidation is close to the value calculated for the maximal possible one-side hydrogen coverage (Fig. 3c) of defective graphene. Further increasing of coverage provides spontaneous dehydrogenation of graphene surface with returning of hydrogen atom to molecular phase.

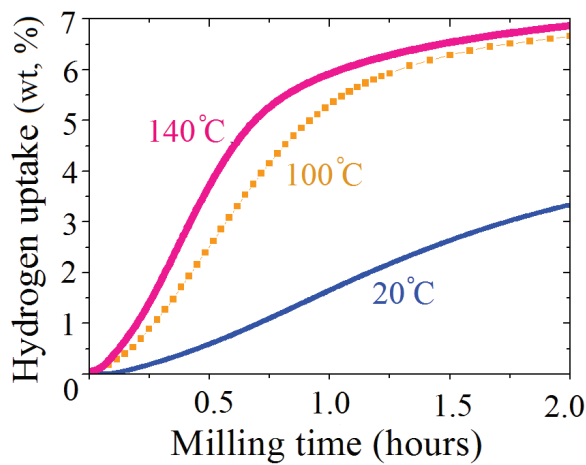

Figure 5. Temperature dependence of hydrogen uptake during Mg hydrogenation in the presence of (5 wt.%) Ni@C nanoparticles

Increasing of the temperature could involve larger carbon area for the hydrogen temporary adsorption reaction. This result corresponds to the experimentally observed increasing of the rate of magnesium hydrogenation at the temperature ~100°C, as compared with room temperature (Fig. 5). Further increase of the reaction temperature till 140℃ does not provide any decay of the reaction time due to depletion of active sites in the vicinity of defects on the surface.

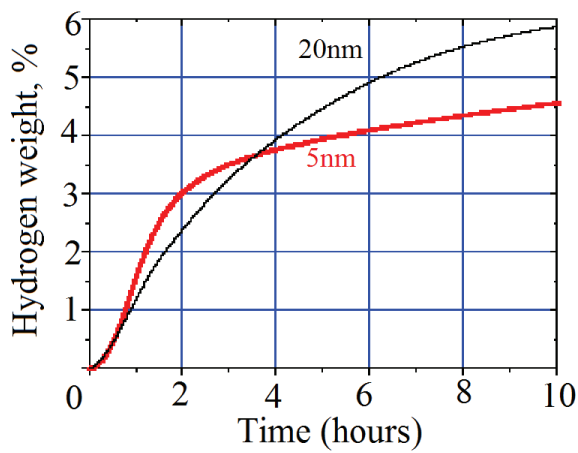

Figure 6. The hydrogen uptake during Mg hydrogenation in the presence of Ni@C of different particle sizes at room temperature.

The last step of our studying of the catalytic properties of Ni@C composites is examination of the effect of size of the nanoparticles. Results of measurements (Fig. 6) demonstrate higher reaction rate of the nanoparticles of smaller size in the first three hours of the catalytic process that could be caused higher curvatures of carbon shells that corresponding with the higher chemical activity of graphene.[30] But this higher chemical activity and larger amount of SW defects caused by higher curvature of carbon shell leads formation of irreversible hydrogenated areas and cessation of Mg hydration process in contrast to the nanoparticles of larger size.

## Conclusion

In summary we reported unconventional catalytic properties of the nickel nanoparticles wrapped in multilayer graphene. Discovered property of this type of nanoparticles to perform the dissociation of molecular hydrogen could be used in the other reactions of dissociation and hydrogenation in the presence of this catalyst. Significant amount of Stone-Wales defects and the presence of nickel substrate provide the transformation of chemically inert multilayer graphene to a different form that is able to generate temporary chemical bond with hydrogen atoms. Reported catalysts based of carbon coated nickel nanoparticles for the hydrogen decomposition can be also used for various reactions. One can wait the similar catalytic properties for graphene on metal substrates.

## Experimental and Theoretical Section

We used a modified gas-condensation installation allowing melting in suspension state and evaporation of liquid metal in a flow of inert gas containing hydrocarbons (e.g., methane or butane). Metal vapor is condensed in a cooler part of a reactor near the evaporation zone; simultaneously hydrocarbon decomposition (pyrolysis) proceeds on the surface of metal nanoparticles. The dimension of formed carbon-coated nanoparticles and a relative ratio of metal and carbon are controlled by varying of the pressure and the flow rate of buffer argon gas and hydrocarbon concentration.

The main characteristics of the process realized by us is that the metallic nucleation centers on which carbon coating occurs are in an inert gas volume, that should result in new morphological features of layered carbon phase. The process of hydrocarbon phase pyrolysis (mainly in butane) excludes oxide and carbide phases formation and can be easily controlled by HREM and Raman spectroscopy. It has been determined that the obtained nanocomposites as a rule consist of spherically-symmetric carbon-encapsulated metal nanoparticles due to a volume distribution of metal clusters and hydrocarbon vapors that are the source of carbon. Carbon layer thickness on a metal nucleus is determined by the concentration of hydrocarbons in a buffer gas.

Our previous studies[5,6] demonstrate that Ni based nanocomposite nanoparticles encapsulated in a carbon matrix have atomic and electronic structure identical to bulk nickel and retain the stability over two years years[6] due to specific properties of the carbon shell. It was shown that such synthesized nanocomposites slightly change their properties when heated in air up to 200 ℃. In the reducing medium these samples have even higher stability. They demonstrated no evidences of metal nucleus oxidation after a biannual storage in the air.

Hydrogenation of magnesium at mechanical treatment was performed in the vibratory ball mill in steel mortars at Mg:balls ratio = 1:140 (on the mass base), magnesium powder loading was 1.5 g. The catalyst (Ni@C, $V_2O_5$, graphite or the mixture of Ni and graphite) was loaded into the same mortar. After purging and filling of mortar by hydrogen the process of grinding was started. The pressure was monitored by pressure controller. Initial hydrogen pressure 850 mm was dropped to 280-380 mm as a result of magnesium hydride formation. The uptake of hydrogen is calculating by the standard formula: Uptake = $\Delta m_{H2}/(m_{Mg} + \Delta_{mH2})$, where $m_{Mg}$ is the mass of load magnesium, and $\Delta mH2$ is the mass of load hydrogen estimated by equation of state of ideal gas in molar form: $\Delta m_{H2} = \Delta PVM/RT$, where $\Delta P$ is the value of decreasing of the pressure, T - temperature of molecular hydrogen with molar mass (M) in known volume (V), and R is the gas constant.

The modelling of molecular hydrogen dissociation has been realized within standard model previously discussed for this reaction over metals and alloys substrate.[30-32] For the understanding of catalytic properties of these nanoparticles density functional theory (DFT) have been performed. We used the pseudopotential code SIESTA,[33] as was done in our previous works.[22,25,28,29] All calculations were performed using the generalized gradient approximation (GGA-PBE)[34] which is feasible for the modeling of hydrogen adsorption and desorption over graphene.[25] Full optimization of the atomic positions was performed. During the optimization, the ion cores were described by norm-conserving pseudo-potentials.[35] The wavefunctions were expanded with a double-ζ plus polarization basis of localized orbitals for carbon and oxygen, and a double-ζ basis for hydrogen. Optimization of the force and total energy was performed with an accuracy of 0.04 eV/Å and 1 meV, respectively. All calculations were carried out with an energy mesh cut-off of 360 Ry and a k-point mesh of 4×4×2 in the Monkhorst-Park scheme.[36] The calculation of chemisorption energies for the hydrogenation was performed using standard equation: $E_{chem} = (E_{host+2nH} - [E_{host} + nE_{H2}])/n$ where $E_{host}$ is the total energy of system before adsorption of 2n atoms of molecular hydrogen with the total energy of molecule – $E_{H2}$. For the modelling of chemisorption the first hydrogen molecule we put pair of hydrogen atoms in orto position on carbon hexagon (see Fig. 3d) or above SW defect (Fig. 3a) according Ref. 22. Connection of the hydrogen atoms with the nearest carbon atoms is necessary for the correct description of dissociation of molecular hydrogen over carbon scaffold. For the modelling of the next steps of chemisorption (Fig. 3b,c and e,f) we examine all possible positions for the addition of the next pair of hydrogen atoms over the nearest pairs of carbon atoms and choose the configuration with the lowest total energy.

## *Acknowledgements*

*DWB acknowledge computational support from the CAC of KIAS. The authors acknowledge the financial support from Russian Foundation of Basic Researches through grants # 10-02-323, 10-03-00372 and 11-03-00820).*

**Keywords:** ((nanoparticles · hydrogen decompositions · graphene · carbocatalysis · first-principles modelling))

[1]  K. Gong, F. Du, Z. Xia, M. Durstock, L. Dai *Science* **2009**, *323*, 760.
[2]  S. Yang, X. Feng, X. Wang, K. Müllen *Angew. Chem. Int. Ed.* **2011**, *50*, 5339.
[3]  S. Wang, D. Yu, L. Dai, D. W. Chang, J.-B. Baek *ACS Nano*, **2011**, *5*, 6202.
[4]  D. R. Dreyer, H.-P. Jia, C. W. Bielawski *Angew. Chem. Int. Ed.* **2010**, *49*, 6813.
[5]  V. R. Galakhov, A. S. Shkvarin, A. S. Semenova, M. A. Uimin, A. A. Mysik, N. N. Schegoleva, A. Ye. Yermakov, E. Z. Kurmaev *J. Phys. Chem. C* **2010**, *114*, 22413.
[6]  V. R. Galakhov, A. Buling, M. Neumann, N. A. Ovechkina, A. S. Shkvarin, A. S. Semenova, M. A. Uimin, A. Ye. Yermakov, E. Z. Kurmaev, O. Y. Vilkov, D. W. Boukhvalov *J. Phys. Chem. C* **2011**, *115*, 24615.
[7]  A. Zaluska, L. Zaluski, J. O. Strom-Olsen *Appl. Phys. A* 2000, *72*, 157.
[8]  L. Schlapbach, A. Zuttel, A. *Nature* **2001**, *414*, 353.
[9]  Y. Chen, J. S. Williams *J. Alloys Comp.* **1995**, *217*, 181.
[10] J. Ryden, B. Hjorvarsson, T. Ericsson, E. Karlsson, A. Krozer, B. Kasemo *J. Less-Common Met.* **1989**, *152*, 295.
[11] S. Cheung, W.-Q. Deng, A. van Duin, W. Goddard *J. Phys. Chem. A* **2005**, 1i09, 851.
[12] W. Oelerich, T. Klassen, R. Borman *J. Alloys Comp.* 2001, *322*, L5.

[13] A. Ye. Yermakov, V. N. Mushnikov, M. A. Uimin, V. S. Gaviko, A. P. Tankeev, A. V. Skripov, A. V. Solonin, A. L. Buzlukov *J. All. Comp.* **2006**, *425*, 367.
[14] H. Imamura, S.Tabata,Y.Takesue,Y. Sakata, S. Kamazaki *Int. J. Hydrogen Energy* **2000**, 2i5, 837.
[15] H. Imamura, N. Sakasai, Y. Kajii *J. Alloys Compd.* **1996**, 23i2, 218.
[16] C. Z. Wu, P. Wang, X. Yao, C. Liu, D. M. Chen, G.Q. Lu, H.M. Cheng, *J. Alloys Compd.* **2006**, *414*, 259.
[17] A. S. Ingason, A. K. Eriksson, S. Olafsson *J. Alloys Compd.* **2007**, *446–447*, 530.
[18] A. J. Stone, D. J. Wales *Chem. Phys. Lett.* **1986**, *128*, 501.
[19] K. S. Kim, Y. Zhao, H. Jang, S. Y. Lee, J. M. Kim, K. S. Kim, J.-H. Ahn, P. Kim, J.-Y. Choi, B. H. Hong *Nature* **2009** *457*, 706.
[20] X. Li, W. Cai, L. Colombo, R. S. Ruoff *Nano Lett.* **2009**, *9*, 4268.
[21] X. Li, W. Cai, J. An, S. Kim, J. Nah, D. Yang, R. Piner, A. Velamakanni, I. Jung, E. Tutuc, S. K. Baneree, L. Colombo, R. S. Ruoff *Science* **2009**, *324*, 5932.
[22] D. W. Boukhvalov, M. I. Katsnelson *Nano Lett.* **2008**, *8*, 4378.
[23] Q. Wang, D. X. Ye, Y. Kawazoe, P. Jena *Phys. Rev. B.* **2012**, *85*, 085404.
[24] Y. J. Chai, Z. Y. Liu, H.N. Gao, Z.Y. Zhao, N. Wang, D. L. Hou *Int. J. Hydrogen Energy* **2011**, *36*, 14484.
[25] D. W. Boukhvalov, M. I. Katsnelson, A. I. Lichtenstein *Phys. Rev. B* **2008**, *77*, 032427
[26] L. Liu, S. Ryu, M. R. Tomasik, E. Stolyarova, N. Jung, M. S. Hybertsen, M. L. Steigerwald, L. E. Brus, G. W. Flynn *Nano Lett.* **2008**, *8*, 1965.
[27] D. R. Lide ed. (2005). CRC Handbook of Chemistry and Physics (86th ed.). Boca Raton (FL): CRC Press.
[28] D. W. Boukhvalov, Y.-W. Son *Nanoscale* **2012**, *4*, 417.
[29] D. W. Boukhvalov, M. I. Katsnelson *J. Chem. Phys. C* **2009**, *13*, 1416.
[30] B. Hammer, M. Schleffler *Phys. Rev. Lett.* **1994**, *73*, 1400.
[31] B. Hammer, M. Schleffler *Phys. Rev. Lett.* **1995**, *74*, 3487.
[32] D. W. Goodman *Acc. Chem. Res.* **1984**, *17*, 194.
[33] J. M. Soler, E. Artacho, J. D. Gale, A. Garsia, J. Junquera, P. Orejon, D. Sanchez-Portal *J. Phys.: Condens. Matter* **2002**, *14*, 2745.
[34] J. P. Perdew, K. Burke, M. Ernzerhof *Phys. Rev. Lett.* **1996**, *77*, 3865.
[35] O. N. Troullier, J. L. Martins *Phys. Rev. B* **1991**, *43*, 1993.
[36] H. J. Monkhorst, J. D. Park *Phys. Rev. B* **1976**, *13*, 5188.